\documentstyle[12pt,epsf]{article}
\newcommand{\GETPS}[2]{} 				
\renewcommand{\GETPS}[2]{\epsfxsize=#1in\epsfbox{#2}}	
\newcommand{\Newpage}{}
\renewcommand{\Newpage}{\newpage}
\def\bitem{\begin{itemize}}
\def\beq{\begin{equation}}
\def\beqar{\begin{eqnarray}}
\def\eitem{\end{itemize}}
\def\eeq{\end{equation}}
\def\eeqar{\end{eqnarray}}
\def\ack{\section*{Acknowledgement}%
  \addtocontents{toc}{\protect\vspace{6pt}}%
  \addcontentsline{toc}{section}{Acknowledgement}}
\def\nuc#1#2{\relax\ifmmode{}^{#1}{\protect\text{#2}}\else${}^{#1}$#2\fi}

\newcommand{\TITLE}[3]{%
\begin{center}{\Large {#1}}\\ \vspace{0.5cm}{#2}\\ 
\vspace{0.3cm}{\it #3}\end{center}\vspace{0.3cm}}%

\newcommand{\bold}[1]{\mbox{\boldmath $#1$}}    
\newcommand{\AMD}{{AMD}}                        
\newcommand{\FMD}{{FMD}}                        
\newcommand{\QMD}{{QMD}}			
\newcommand{\Z}{\bold{Z}}
\newcommand{\del}{\partial}
\newcommand{\z}{\bold{z}}

\newcommand{\ZC}{\bar{\bold{Z}}}
\newcommand{\r}{\bold{r}}
\newcommand{\p}{\bold{p}}
\newcommand{\VEV}[1]{\langle{#1}\rangle} 
\newcommand{\SEV}[1]{\prec{#1}\succ}	

\newcommand{\ket}[1]{|{#1}\rangle}
\newcommand{\Hml}{{\cal H}}     
\newcommand{\Wei}{{\cal W}}     
\newcommand{\E}{{\scriptscriptstyle{E}}}
\newcommand{\comment}[1]{{}}

\newcommand{\FIGCAPW}[1]{%
\begin{center}\begin{minipage}{14cm}\begin{small}{#1}\end{small}\end{minipage}\end{center}}
\def\FIGcanl{\begin{figure}
\GETPS{6}{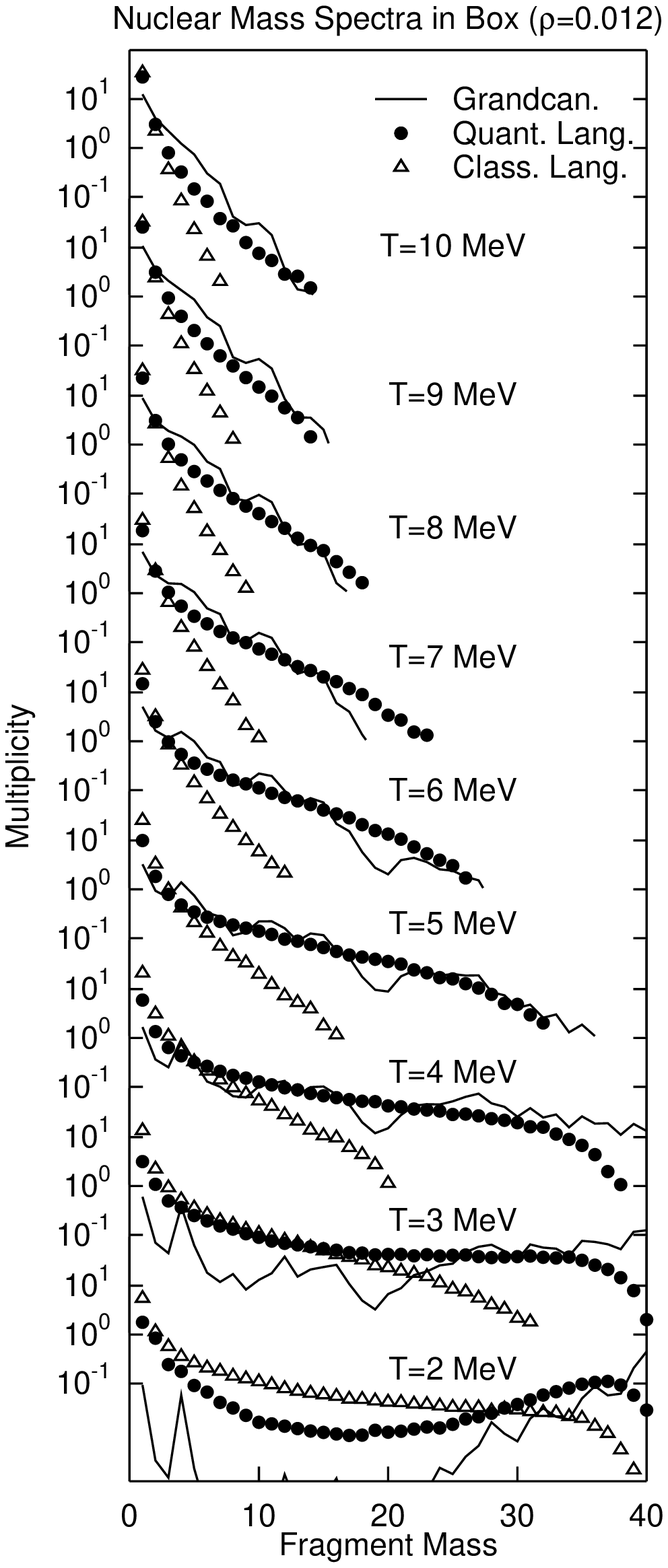}
\caption{Mass distribution in a canonical ensemble of 40 nucleons.}
\label{fig:canl}
\FIGCAPW{
	Circles and triangles show the calculated results of \QMD\
	at the indicated temperatures $T$
	with ($\alpha<1$) and without ($\alpha=1$) the quantal modification.
	As an instructive reference for the quantal calculation,
	we also show the grand canonical results (solid lines)
	based on a simple statistical treatment \cite{grand}.
}
\end{figure}}
\def\FIGimfb{\begin{figure}
\begin{center}
\GETPS{6}{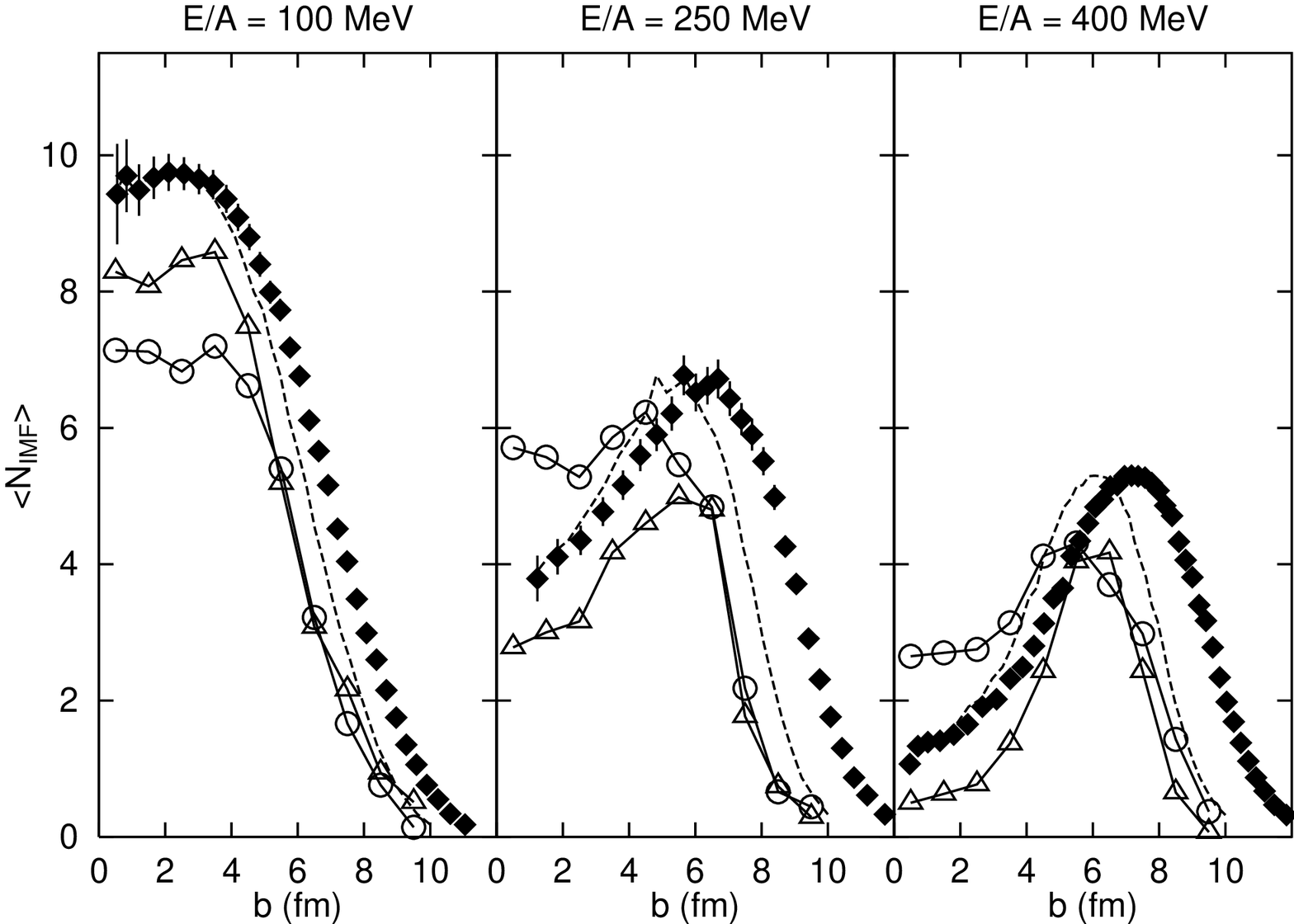}
\caption{IMF multiplicity before statistical decay.}
\label{fig:imfb}
\FIGCAPW{
	Circles and triangles indicate \QMD\ results at given energies
	with and without the quantal Langevin force, respectively.
	The experimental data~\cite{Tsang} are shown by solid diamonds.
	Dotted lines show the experimental data using a scaled impact parameter
	assuming a maximum impact parameter of 10 fm.
	The detector efficiency is not taken into account in the calculation.
}
\end{center}
\end{figure}}
\def\FIGimfa{\begin{figure}
\begin{center}
\GETPS{6}{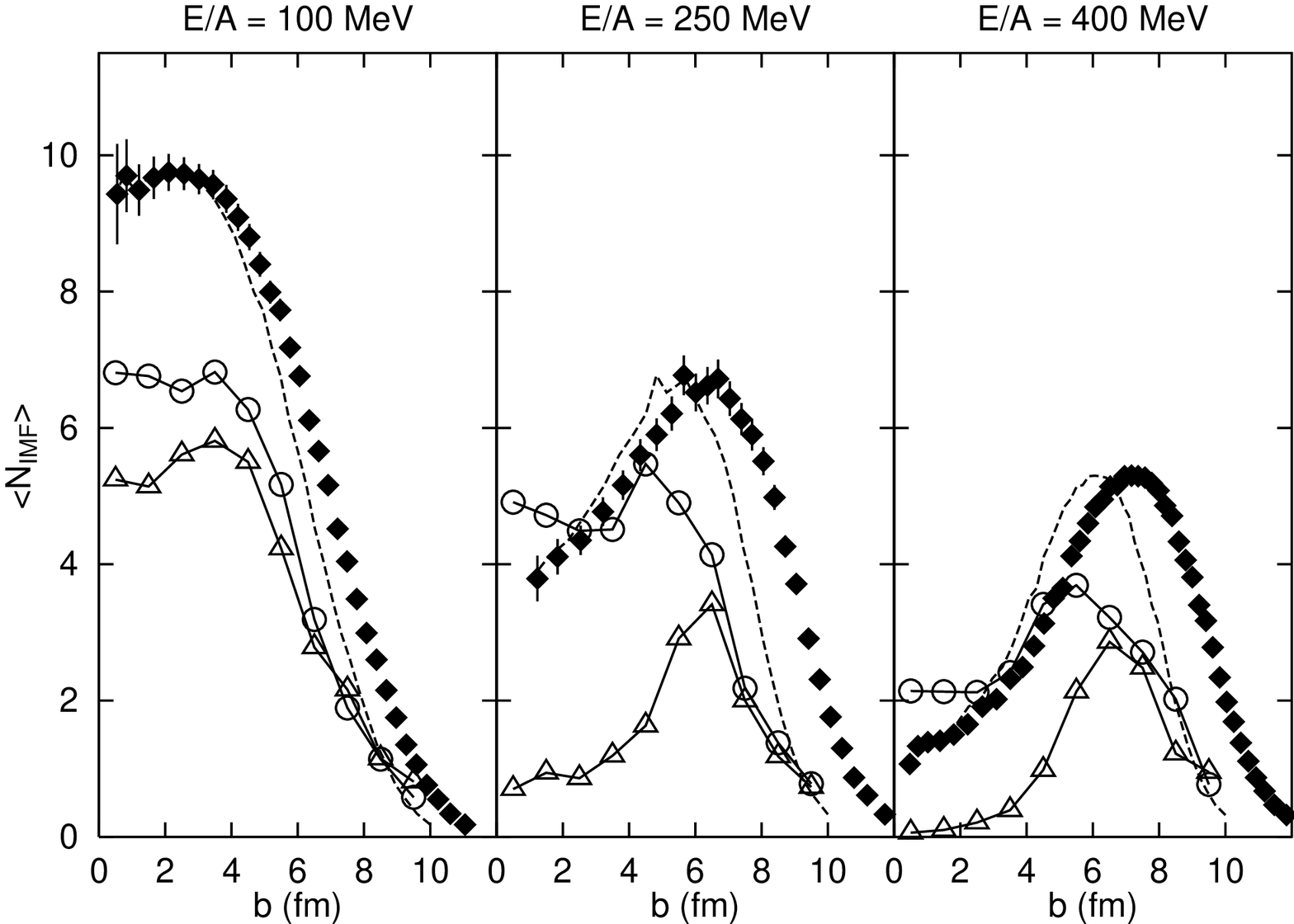}
\caption{IMF multiplicity after statistical decay.}
\label{fig:imfa}
\FIGCAPW{
	The display is similar to fig.\ \ref{fig:imfb}.
	Circles and triangles show the results of the combined \QMD\ and 
	statistical decay model (SDM)~\cite{SDM} calculation
	with and without the quantal Langevin force, respectively.
}
\end{center}
\end{figure}}
\def\FIGimft{\begin{figure}
\begin{center}
\GETPS{6}{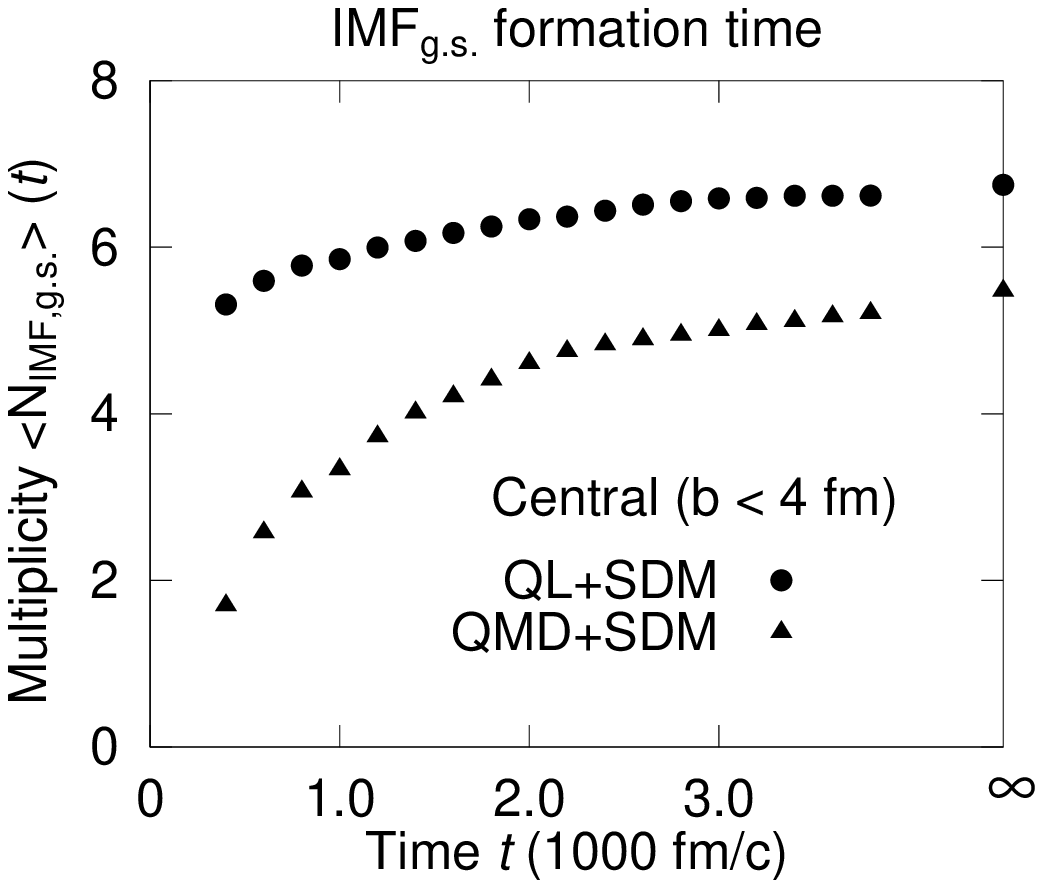}
\caption{Formation time of bound IMFs.}
\label{fig:imft}
\FIGCAPW{
	Circles and triangles indicate the time dependence
	of the mean number of bound intermediate-mass fragments,
	as a result of the combined \QMD\
	and statistical decay model (SDM)~\cite{SDM} calculation
	with and without the quantal Langevin force, respectively,
	for central collisions ($b<4$ fm).
}
\end{center}
\end{figure}}
\begin{document}
\begin{titlepage}
\noindent{\sl Physics Letters B}	\hfill LBNL-39473, HUPS-96-5\\[4ex]

\TITLE{
        Quantum Fluctuation Effects on Nuclear Fragment Formation$^*$
}{
        Akira Ohnishi${}^{a,b}$ and J\o rgen Randrup$^{b}$
}{
        $a.$\ \ Department of Physics, Faculty of Science\\
                Hokkaido University, Sapporo 060, Japan\\[0.5ex]
        $b.$\ \ Nuclear Science Division,
                Lawrence Berkeley National Laboratory\\
                University of California, Berkeley, California 94720, USA
}

\begin{center}
October 31, 1996	
\end{center}
~\\[2ex]
\begin{abstract}
Multifragmentation in Au+Au collisions is investigated 
at incident energies in the range 100-400 MeV per nucleon
by means of a recently developed quantal Langevin model.
The inclusion of quantum fluctuations
enhances the average multiplicity of intermediate mass fragments,
especially in central collisions.
This is mainly because the excitation energies of fragments 
are reduced due to the quantal behavior of intrinsic specific heat.
\end{abstract}

\vfill\noindent$^*$
This work was supported in part by
the Grant-in-Aid for Scientific Research (No.\ 06740193)
from the Ministry of Education, Science and Culture, Japan,
and by the Director, Office of Energy Research,
Office of High Energy and Nuclear Physics,
Nuclear Physics Division of the U.S. Department of Energy
under Contract No.\ DE-AC03-76SF00098.

\end{titlepage}

\section{Introduction}\label{sec:Intro}

One of the main goals of heavy-ion physics is to explore the
various phases of nuclear matter
and achieve an understanding of the associated phase transitions.
At intermediate energies,
the nuclear liquid-gas phase transition is of primary interest,
and the efforts to understand this phenomenon have intensified
following the recent extraction of a caloric curve
of hot nuclear matter~\cite{LGPT}
suggesting that the phase transition is of first-order
and occurs at a temperature of 4$-$5 MeV.
The first-order nature of the phase transition
may lead to an enhancement of fragment formation~\cite{Texas,GSI,Tsang},
since the rapid expansion following an energetic nucleus-nucleus collision
causes a transformation from a hot to a supercooled gas phase.
The resulting dilute system is mechanically unstable and 
fragments are therefore likely to be formed,
in a manner similar to the formation of water drops.

The experimental exploration of nuclear phase transitions
must rely on nuclear collisions,
and it is therefore necessary to invoke dynamical models
in the interpretation of the data.
In addition to the complications arising from the non-equilibrium
character of the collision dynamics,
it is also important to ensure that the models employed
take due account of the quantal nature of the nuclear system.

In order to illustrate this important point,
we note that the nuclear liquid-gas phase transition differs significantly
from the usual liquid-gas phase transition in macroscopic matter.
The largest difference lies in the role played by quantal statistics.
In the case of usual macroscopic matter,
the total energies are to a good approximation
linear functions of the temperature in both the liquid and gas phases.
Thus the effective number of degrees of freedom is essentially constant
in each phase.
In contrast to this familiar situation,
the liquid phase of a nucleus exhibits an increase in the number of 
activated degrees of freedom as the temperature is raised.
For example, the excitation energy of a nucleus at low temperature
increases like $E^* = a T^2$
(where the level density parameter is $a\approx A/(8 {\rm MeV})$),
which is a typical quantal behavior,
while the gas phase is characterized by the usual classical relation
$E^*/A=3T/2$.
The two curves intersect at $T\approx12$ MeV,
which is much higher than the suggested transition temperature.
This indicates that the quantal statistical nature of the nuclear system
plays an important role for the phase transition and, presumably,
for the associated nuclear multifragmentation processes.

To take approximate account of the quantal features
of the evolving nuclear system,
we augment the socalled Quantum Molecular Dynamics (\QMD)
model~\cite{QMD,Boal,Maruyama}
by a stochastic term given by the recently developed
quantal Langevin model~\cite{OR95,OR96a}.
The \QMD\ model describes the nuclear many-body system
in terms of a product wave-function of gaussian wavepackets
for the individual nucleons and has been successful
in accounting for a variety of collision observables,
such as collective flow and particle production \cite{QMD}.
The \QMD\ model and its antisymmetrized versions,
Antisymmetrized Molecular Dynamics (\AMD)~\cite{AMD}
and Fermionic Molecular Dynamics (\FMD)~\cite{FMD},
have also met with some success in describing fragment production
\cite{QMD,Maruyama,AMD,Nara,NMD,Iwamoto}.
The fragments formed in these molecular dynamics calculations
are generally far too excited to remain intact,
so the subsequent decay must be accounted for
in order to extract the yield of bound fragments
that are measured experimentally.
This combination of molecular dynamics and subsequent (usually statistical)
decay calculation may give us a correct picture of fragment formation
when the source of fragments is limited
to one compound nucleus or excited projectile-like and target-like fragments.
However, when the copious intermediate-mass fragments (IMF) appear
simultaneously, 
the above fragment formation mechanism becomes dubious.
For example, 
although the calculated IMF multiplicity without the subsequent decay
seems to reproduce the data reasonably~\cite{NMD,Iwamoto}, 
the decay processes largely eliminates the IMF yield~\cite{Iwamoto}.

We suggest that part of the reason for this consistent shortcoming
of the entire class of \QMD-like models
may be found in the fact that the model is effectively classical,
since the equations of motion for the wavepacket parameters
have been derived by a standard variational principle.
The resulting description is then the dynamics of the wavepacket centroids
and the inherent quantum fluctuations are neglected.
The presence of quantum fluctuations is signaled by the fact that
the wavepackets are not energy eigenstates.
This basic feature generally causes the expectation value of the Hamiltonian
to fluctuate in the course of time.

In order to remedy the situation,
a quantal Langevin model was recently developed \cite{OR95,OR96a}.
The Langevin force enables the wavepacket system
to explore its entire energy spectral distribution,
rather than being restricted to its average value,
and it leads to a much improved description of the quantum statistical features.
In particular,
the resulting specific heat now exhibits the characteristic
evolution from a quantum fluid towards a classical gas
as a function of temperature~\cite{OR95,OR96a},
in contrast to the behavior emerging with the usual treatment.
Since a change of a fragment's specific heat
is associated with a change in its statistical weight,
the effect is clearly relevant for the fragment production problem.

The key features of the results obtained with the quantal Langevin model
are the occurrence of larger fluctuations
and an enhancement of stable configurations,
such as bound fragments,
as a result of the need to take account of the spectral distortion
of the wavepackets.
The former feature arises from the fact that
the wave packets are populated according to
the strength of the eigencomponent for the given energy,
and therefore the wavepacket parameters can have larger fluctuations
than when the expectation value of the energy is conserved.
On the other hand,
in order to project out the appropriate energy component from the wavepacket,
it is necessary to take account of its internal distortion.
The combination of these two basic features is then expected to enhance
the IMF yield at the final stage:
while the larger fluctuations allow the system to explore
more configurations and thus enhances the yield of primary fragments,
the latter stabilizes the fragments,
since the compensation for the quantum distortion
effectively acts as a cooling mechanism.

In our previous work \cite{OR96a}, 
we have applied the quantal Langevin model to a simple soluble example
and have shown that it leads to the correct quantal microcanonical equilibrium.
However,
in order to apply this model to actual nuclear collisions, 
the treatment of the center-of-mass cluster motion must be suitably modified.
In the \QMD\ and \AMD\ treatments,
the single-particle wave function is represented by a gaussian wave packet
with a fixed width.
Then the center-of-mass wave function of each cluster is also a gaussian
wave packet and, accordingly,
it has zero-point kinetic energy
as well as an energy fluctuation dependent on the velocity.
Within the framework of the quantal Langevin model,
this energy fluctuation causes
a quantal fluctuation of the CM motion of the cluster 
even if it is well separated from other clusters and nucleons.
This phenomenon reflects the expansion of a localized free wave packet
and should be eliminated,
as is usually done in Hartree-Fock and \AMD\ calculations, for example,
before the binding energy is calculated.
In order not to modify the $Q$-values in the fragmentation process,
the cluster CM motion must also be removed in the dynamical evolution,
as was already discussed in refs.~\cite{AMD,eQMD}.
Therefore, we have removed the energy fluctuation
of the cluster center-of-mass motions as well.
This can be accomplished by considering
the motion of the individual nucleon constituents
relative to to the local collective velocity.

In this letter, we have included the quantal Langevin force into
the framework of \QMD.
Although we have formulated the quantal Langevin model 
within the \AMD\ framework,
the appearance of the quantal fluctuation originates in the wave-packet 
nature of the many-body state and is therefore a general phenomenon.
Thus the basic features of the model also apply to \QMD\
which is easier to apply to heavier systems
and we focus our attention on the fragmentation process
in \nuc{197}{Au}+\nuc{197}{Au} collisions.
In this system, the total number of nucleons is so large
that the statistical properties are expected to play a major role.
In addition, 
the multiplicities of intermediate mass fragments are measured
at incident energies of 100, 250, and 400 MeV per nucleon and their impact
parameter dependence is also deduced~\cite{Tsang}.
The IMF multiplicities reach up to around ten,
which is the largest number observed so far.

\section{Quantal Langevin model at fixed temperature}\label{sec:QLT}

We first give a condensed description of the recently introduced
quantal Langevin model for the situation when the system can be
regarded as being in thermal equilibrium at a given temperature.

The treatment seeks to take account of the energy fluctuations present
in a system being described in terms of many-body wave packets.
As we have already discussed in detail in Ref.\ \cite{OR96a}, 
this inherent energy dispersion modifies the statistical weight 
relative to the naive the classical form,
\beq
\Wei_\beta(\Z)\ \equiv\ \VEV{\Z|\exp(-\beta\hat H)|\Z}\
	\not=\ \exp(-\beta\Hml)\ .
\eeq
Here $\Hml = \VEV{\Z|\hat H|\Z}$ is the expectation value of the Hamiltonian
in the given wave-packet state $\ket{\Z}$ and thus the last quantity represent
the usual classical statistical weight.
The complex parameter set $\Z=\{\z_1,\z_2,\ldots \z_A\}$ 
is related to the phase space coordinates in QMD,
$\z_n ={\r_n/2\Delta r + i\p_n/2\Delta p}$,
where $\Delta r$ and $\Delta p$ are widths of wave packet.
By invoking the harmonic approximation,
it is possible to obtain a good description of
the statistical weight by means of a simple ``free energy",
\beq
\label{F_q}
{\cal F}_\beta(\Z)
	\equiv -\log \Wei_\beta(\Z)
	= {\Hml\over D} \left( 1-\exp(-\beta D) \right)
	\ ,
\eeq
where $D=\sigma_\E^2/E^*$ is the effective level spacing.
(The energy of the wave packet relative to its ground state is denoted by
$E^*$ and $\sigma_\E^2$ is the associated variance.)

The relaxation towards this approximate quantal equilibrium
can be described by the following Fokker-Planck equation
for the distribution $\phi(\Z)$ of wave-packet parameters,
\beqar
\label{FP}
{D\phi \over Dt}\ 
&=&\ 
	\left[
		-\sum_i {\del\over\del q_i} V_i
		+\ \sum_{ij} {\del\over\del q_i} M_{ij} {\del\over\del q_{j}}
	\right] \phi\ , \\
V_i\ 
	&=&\  - \sum_{j} M_{ij} {\del{\cal F}_\beta\over\del q_j}\ , 
\eeqar
where $q_i$ represents either $\r_n$ or $\p_n$.
It is easy to check that
the statistical equilibrium distribution, $\phi_{\rm eq}=\exp(-{\cal F}_\beta)$,
is a stationary solution to the above Fokker-Planck equation.
Moreover, when the classical statistical weight is employed
({\em i.e.} when ${\cal F}^C_\beta=\beta\Hml$),
the drift and the diffusion coefficients of the Fokker-Planck equation
satisfy the usual Einstein relation,
\beq
\label{EinsteinC}
V^C_i = - \beta \sum_{j} M_{ij}{\del\Hml \over \del q_j}\ .
\eeq
On the other hand, when the quantal statistical weight
obtained with the harmonic approximation is used, eq.\ (\ref{F_q}),
the relation is modified.
For example, if the effective level spacing
$D$ does not depend strongly on the wave-packet parameters,
the drift coefficient reduced by the factor $\alpha$,
\beq
\label{EinsteinQ}
V_i = - \alpha\beta \sum_{j} M_{ij}{\del\Hml \over \del q_j}\ ,
	\quad
\alpha= {1-\exp(-\beta D) \over \beta D} \ .
\eeq
Since $\alpha$ is smaller than unity,
the resultant Fokker-Planck equation gives smaller friction,
thus in effect relatively larger fluctuations will arise.

It is convenient to solve the Fokker-Planck transport equation
by means of a Langevin method.
Within the framework of \QMD\ the Langevin equation becomes
\beqar
\label{QLp}
\dot{\p} &=&	
	\bold{f}-\alpha\beta\bold{M}^p\cdot\bold{v}
	+\bold{g}^p\cdot\bold{\xi}^p \ , \\
\label{QLr}
\dot{\r} &=&	
	\bold{v}+\alpha\beta\bold{M}^r\cdot\bold{f}
	+\bold{g}^r\cdot\bold{\xi}^r \ ,\\
\bold{v} &=& {\del\Hml\over\del\p}\ ,\quad
\bold{f} =-{\del\Hml\over\del\r}\ ,
\eeqar
\beq
\bold{M}^p =
	 \bold{g}^p\cdot\bold{g}^p \ , \quad
\bold{M}^r =
	 \bold{g}^r\cdot\bold{g}^r \ .
\eeq
Here $\r$ and $\p$ are the phase-space centroid parameters for the
wave packet and $\xi$ is used to denote random numbers drawn from a
normal distribution with two a variance equal to two.
In these equations,
we have omitted the diffusion-induced drift term
and that part of the mobility tensor that connects $\r$ and $\p$.

\subsection{Energy Fluctuation}

The key quantity in the quantal Langevin treatment is
the energy dispersion $\sigma_\E^2$ within each wave packet.
In Refs.\cite{OR95,OR96a},
the following approximate form was derived,
$\sigma_\E^2$.
\beq
\sigma_\E^2
	= {\del \Hml \over \del \Z}\cdot \bold{C}^{-1}\cdot
		{\del \Hml \over \del \ZC}\
	=\sum_n	\left[
		\Delta r^2\ \bold{f}_n\cdot\bold{f}_n
		+\Delta p^2\ \bold{v}_n\cdot\bold{v}_n
		\right] \ .
\eeq
Here the matrix $\bold{C}$ arises from the antisymmetrization and,
accordingly, it is the unit matrix in \QMD.
There are two problems in the application of the above expression 
to dynamical processes within the framework of QMD,
such as nucleus-nucleus collisions. 
The first concerns the spurious zero-point center-of-mass motion of clusters
and the second arises from the Pauli blocking,
as we shall now discuss.

The energy fluctuation becomes larger when the wave packet moves faster,
as should be the case,
and it results from the zero-point kinetic energy of the
fragment center of mass. 
The expression is adequate when the configuration volume is compact
and all the degrees of freedom can be treated as wave packets.
However,
as is already discussed in Refs.\ \cite{AMD,eQMD}, 
this zero-point CM kinetic energy is spurious
and should be removed from the Hamiltonian.
It must then also be removed from the energy fluctuation.
In addition, 
it may be unreasonable to carry out the integral over all the parameter space 
near the given value,
since no Pauli blocking effect is taken account of 
in the phase space parameters themselves in QMD.
Then, it appears more physical to reduce the energy fluctuation
according to the appropriate Pauli blocking factor.
In order to satisfy these two requirements
we have adopted the following form of the energy fluctuation,
\beqar
\sigma_\E^2
	&=&\sum_n\ 
		(1-f_n)^2\ 
		\left[
		\Delta r^2\ \bold{f}_n\cdot\bold{f}_n
	+	\Delta p^2\ (\bold{v}_n-\bold{u}_n)
			\cdot(\bold{v}_n-\bold{u}_n)
		\right] \ . \\
\bold{u}_i
	&=&	{1\over N_i}\sum_j c_{ij}\, \bold{v}_j
	\ .
\eeqar
The additional factor $(1-f_n)^2$ represents the Pauli blocking
for small changes,
where $f_n$ is the Wigner function.
Here $\bold{v}_j$ denotes the velocity of the nucleon $j$
and the local flow velocity $\bold{u}_i$ is obtained by performing
a suitable smoothing,
using $c_{ij}=\exp(-r_{ij}^2/4\Delta r^2)$ with $N_i=\sum_jc_{ij}$.
The quantity entering in the energy fluctuation is then $\bold{v}-\bold{u}$,
the velocity of a wave packet relative to the local collective flow,
which represents those degrees of freedom having a wave packet nature.

On the other hand, the collective velocity $\bold{u}$,
which is thought of as the cluster velocity,
represents the degrees of freedom having a classical nature.
The drift term of eq.\ (\ref{QLp}) is therefore modified accordingly,
\beq
\label{QLpCM}
\dot{\p} =	
	\bold{f}-\alpha\beta\bold{M}^p\cdot(\bold{v}-\bold{u})\ 
	-\beta\bold{M}^p\cdot\bold{u}\ 
	+\bold{g}^p\cdot\bold{\xi}^p \ .
\eeq
It should be noted that when the energy fluctuation $\sigma_\E^2$ is small,
the above Langevin equation leads to the normal (classical) Langevin 
equation, 
$\dot{\p} =	
	\bold{f}-\beta\bold{M}^p\cdot\bold{v}+\bold{g}^p\cdot\bold{\xi}^p$.
This situation is realized when all the fragments are well separated
and the intrinsic energy fluctuation is very small
({\em i.e.} all the fragments are close to their ground state).

\subsection{Thermal distortion and observation}

In addition to modifying the statistical weight,
the energy fluctuation also modifies the meaning of wave packet ensemble,
since it causes a thermal distortion of the spectral strength distribution
of the energy eigencomponents within each wave packet.
It is important to take account of this effect.

Since the expectation value of an observable $\cal O$ is generally given by
\beq
\SEV{{\cal O}}\
	=\ {{\rm Tr}\left(\hat{O}\exp(-\beta\hat{H})\right)
	 \over {\rm Tr}\left(\exp(-\beta\hat{H})\right)}\ .
\eeq
In order to obtain this quantity as an weighted average over the thermal
ensemble of wave packets,
with the weight factor being the usual statistical weight $\Wei_\beta(\Z)$,
it is necessary to distort the individual wave packets appropriately.
The resulting contribution to $\SEV{{\cal O}}$ from a given wave packet
is then temperature dependent,
\beq
{\cal O}_\beta(\Z)
	=	{1\over \ \Wei_\beta(\Z)}\
		\VEV{\Z|\exp(-\beta\hat H/2)\ \hat O\ \exp(-\beta\hat H/2)|\Z}
		\ .
\eeq

The distortion operator $\exp(-\beta\hat H/2)$ reduces the expectation value 
of the Hamiltonian in the particular state $\ket{\Z}$.
The thermal distortion is calculated by replacing the time $t$
by the imaginary time $i\tau$ in the equation of motion.
The resulting ``evolution'' is then described by a cooling equation,
\beq
{d\p_n \over d\tau}=-{2\Delta p^2\over\hbar}\,(\bold{v}_n-\bold{u}_n) \ ,\quad
{d\r_n \over d\tau}= {2\Delta r^2\over\hbar}\,\bold{f}_n		\ ,
\eeq
with which the state should be propagated until $\tau=\hbar\beta/2$.
Here, $\bold{v}$ is again replaced by $\bold{v}-\bold{u}$
in order to leave the collective (or cluster) motion unaffected.

\subsection{Fragmentation at fixed temperature}

We have incorporated the above described quantal Langevin force 
into the \QMD\ model.
In the present implementation,
the Gogny interaction~\cite{Gogny} is used as the effective interaction.
In addition, a correction for the fragment zero-point kinetic energy
is included as described in refs.~\cite{AMD,eQMD}.
Moreover,
since the quantal Langevin force is sensitive to
the minimum state of the given Hamiltonian,
a Pauli potential \cite{eQMD,Ohnishi92a}
to guarantee that the nuclear sizes and bindings
are reasonably reproduced by the model.

Figure~\ref{fig:canl}
shows the resulting mass distribution of fragments
as obtained at various temperatures having a low average density
(to ensure that the fragments are well separated).
The results of the quantal Langevin model agree well
with  the grandcanonical calculation when $T \geq 4$ MeV.
At lower temperatures,
probably because of the limited functional space of \QMD, 
both the quantal Langevin model and the classical Langevin model
fail to describe the population of the heaviest fragments
which in turn may lead to the overestimation of the proton multiplicity.

\FIGcanl

It is interesting to note that the mass distribution changes its appearance
in the temperature region $T= 4\sim 6$ MeV.
At lower temperatures,
a large part of the nucleons are bound in large fragments,
while at higher temperatures, the lighter fragments grow abundant.
This behavior reflects the transition from
the excitation of various collective modes of heavy nuclei 
to the realization of fragmentation degrees of freedom.
In our previous work~\cite{OR95,OR96a},
we have shown that this feature can be also seen from 
the relationship between temperature and excitation energy.
The maximum specific heat is obtained for the temperature $T\sim 4$ MeV,
primarily because the fragment degrees of freedom
are activated rapidly as $T$ is increased above this value.

\section{Application to nuclear collisions}
\label{sec:QLE}

In order to apply the treatment described above
to actual nucleus-nucleus collision processes,
some important modifications of the \QMD\ model are required.
In the usual \QMD\ treatment,
the expectation value of the Hamiltonian is a constant of motion
(since the equations of motion for the wavepacket parameters
have been derived from a standard variational principle).
When the quantum fluctuations are taken into account,
the inherent energy fluctuation within each many-body wavepacket
makes it necessary to include wavepackets having different
energy expectation values.
At first, this feature might seem unreasonable
as it appears to violate energy conservation.
However,
an energy eigenstate can be constructed as a superposition
of wavepackets whose energy expectation values
(the diagonal matrix elements of the Hamiltonian operator)
are not necessarily the same as the energy eigenvalues.
This feature grows more prominent in the reaction region
and there is therefore no reason to exclude wavepackets
with different energy expectation values in the dynamics, 
especially during the strongly interacting stage.

\subsection{Quantal Fokker-Planck equation at a fixed energy}

By proceeding as in the case considered above where the temperature is given,
it is possible to derive a quantal Fokker-Planck equation
for situations in which the energy is specified,
such as the evolution of an isolated system.
The equilibrium distribution is now a microcanonical ensemble,
with a corresponding statistical weight for each wave packet, 
\beq
\rho_\E(\Z)\ \equiv\ \VEV{\Z|\delta(E-\hat H)|\Z}\
	\propto\ {(\Hml/D)^{E/D}\over \Gamma(E/D+1)} \exp(-\Hml/D)	\ .
\eeq
The last relation assumes that the spectral function is
given by a continuous Poisson distribution.
The drift coefficient can then be readily calculated,
\beqar
V_i\ 
	&=&\  - \sum_{j} M_{ij} {\del{\cal F}_\E\over\del q_j}\
	\simeq\  - \beta_\Hml \sum_{j} M_{ij} {\del\Hml\over\del q_j}	\ ,\\
{\cal F}_\E(\Z)
	&\equiv&	-\log\rho_\E(\Z)
	\ ,\\
\beta_\Hml
	&\equiv&	{\del{\cal F}_\E\over\del\Hml}\
	=\		{\Hml - E \over \sigma_\E^2}\ .
\eeqar
It is interesting to note that the drift term acts to restore the energy:
 when the Hamiltonian $\Hml$ is greater than the given energy $E$,
the drift term reduces the Hamiltonian
(since $\beta_\Hml$ is positive in this case and the mobility tensor is 
always positive definite.
Furthermore, since the phase volume with larger $\Hml$ is larger,
the dynamical trajectory usually goes through the region with $\Hml>E$.
Therefore,
the drift term generally reduces the excitation energy of fragments
in the final stage of collision.
In the asymptotic region where all the fragments are well separated
and all of them are sufficiently cold,
the energy fluctuation becomes small
and the energy expectation value converges to the given value.

Thus,
the basic features of the quantal Langevin model remain the same
also in the case of a fixed energy,
namely larger fluctuations and intrinsic distortion.
To show this important point,
we introduce the state-dependent microcanonical inverse temperature $\beta_\E$,
whose average over the entire phase space with the weight $\rho_\E(\Z)$
gives the global microcanonical temperature,
\beq
\beta_\E(\Z)\
	\equiv\	{\del \over \del E}\ \log\rho_\E(\Z)\
	=\	-{\del \over \del E}\ {\cal F}_\E(\Z)\
	\simeq\	{1\over D}\ \log(\Hml/E)\ .
\eeq
The last relation holds when we employ the continuous Poisson distribution
and the effective level spacing $D$ is small compared to $\Hml$ and $E$.
In terms of this $\beta_\E$, we have 
$
	\beta_\Hml = \alpha_\E\ \beta_\E \ ,
$
with $\alpha_\E= ( 1 - \exp(-\beta_\E D))/\beta_\E D$. 
Thus the drift term is reduced by the factor $\alpha_\E$ compared
to the normal Einstein relation,
in analogy with the canonical case of a fixed temperature.

\subsection{Mobility tensor}

Although the preceding statistical discussion gives the form of 
Fokker-Planck equation,
including the modified fluctuation-dissipation relation,
the mobility tensor must be determined separately.
Although the mobility tensor does not affect the equilibrium properties,
it is important in a dynamical scenario
since it determines the time scale.
As we discussed in refs.~\cite{OR95,OR96a}, 
this mobility tensor essentially represents the transition rate
from the current dynamical state.
This transition rate should be described by the residual interaction
between nucleons,
{\em i.e.} the off-diagonal part of the original Hamiltonian operator.
In the present study,
we have used the following relatively simple mobility tensor,
\beqar
M^r_{ni,mj}
	&=& g\ \delta_{nm}\delta_{ij}\ \Delta r^2\ {\sigma^r_n \over \hbar}\ 
		(1-f_n) \ ,\\
M^p_{ni,mj}
	&=&g\ \delta_{nm}\delta_{ij}\  \Delta p^2\ {\sigma^p_n \over \hbar}\ 
		(1-f_n) \ ,\\
(\sigma^r_n)^2
	&=&	\Delta r^2\ \bold{f}_n\cdot\bold{f}_n
			\ ,\\
(\sigma^p_n)^2
	&=&	\Delta p^2\ (\bold{v}_n-\bold{u}_n)
			\cdot(\bold{v}_n-\bold{u}_n)
			\ .
\eeqar
Since the transition induced by the quantal Langevin force
is of one-body nature,
we have incorporated the blocking factor $1-f_n$.

\subsection{Results for Au+Au collisions}\label{sec:Results}

We have applied the quantal Langevin model to collisions of two gold nuclei
at incident energies of 100, 250, and 400 MeV per nucleon.
In addition to the Hamiltonian discussed above,
an energy-conserving two-body collision term is also included.
The average multiplicity of intermediate-mass fragments
calculated for \nuc{197}{Au}+\nuc{197}{Au} is shown in fig.~\ref{fig:imfb}.
The treatments with and without the quantal Langevin force
agree with each other qualitatively.
For example,
the IMF multiplicity has the peak at central collisions at $E/A=100$ MeV,
while the peak moves to finite impact parameter at higher incident energies.

\FIGimfb

\Newpage
Since the fragments extracted from the dynamical calculations are still excited,
it is necessary to include the subsequent statistical decay
before contact with experiment can be made.
Figure~\ref{fig:imfa} shows the resulting average IMF multiplicity
after the statistical decay chains have deexcited the fragments to
below the light-fragment separation energy
(we shall consider such relatively cold fragments
as being in their ground states).
As the statistical decay model (SDM), we have used the evaporation
model of P\"uhlhofer~\cite{SDM}.
In the case of \QMD\ without quantal Langevin force,
the excitation energies of the primary fragments
are large enough to largely eliminate the IMF component from the
final mass distribution.
This tendency is more clear for central collisions,
where the fragments are likely to be highly excited.
On the other hand, \QMD\ with the quantal Langevin force
reduces the excitation energies of the primary fragments
because of the quantum-statistical nature of the intrinsic degrees of freedom.
As a result,
the difference between the average multiplicities before and after the
statistical decay is less than one in the quantal Langevin model.

\FIGimfa

\Newpage
Although the difference between the two models seems small at $E/A=100$ MeV,
the IMF formation mechanisms are quite different.
This feature becomes more clear when the ground state population is examined.
With the quantal Langevin force,
the primary fragments emerge relatively cold as a results of the drift term
(the intrinsic distortion referred to above),
while there is no other way than nucleon emission to cool
fragments in the normal QMD model.
Thus,
the ground-state IMFs are more easily formed in the quantal Langevin model.
The evolution of the multiplicity through the statistical decay stage
is shown in fig.~\ref{fig:imft}.
It is clear that a large part of the stable IMFs are created in the
dynamical stage in the quantal Langevin model,
whereas almost all the IMFs that appear with the usual QMD treatment
owe their stability to the emission of particles
during in the statistical decay stage.
Such a difference can, in principle, be subjected to measurement
by exploiting the momentum correlation of the fragments.

\FIGimft

\section{Concluding remarks}\label{sec:summary}

In this paper,
we have adapted the recently developed quantal Langevin model
to simulations of nuclear collisions
by embedding it into the widely used Quantum Molecular Dynamics model
in which the individual nucleons are described by gaussian wave packets.
The essential modification arises from the fact that the nuclear collision
proceeds in isolation,
so it is appropriate to consider the energy rather than the temperature
to be fixed.

The essential effect of incorporating the quantal Langevin model
is that the system exhibits larger fluctuations
and the degree of excitation of the emerging fragments is smaller.
These two features conspire to enhance the production
of intermediate-mass fragments.

The usual molecular-dynamics treatment leads to primary fragments
that are typically sufficiently excited to emit nucleons.
It is therefore essential to add an ``afterburner''
that subjects the unstable fragment to statistical decay.
This process causes a strong suppression of the IMFs,
and a corresponding enhancement of lighter fragments.
Therefore,
even though the usual dynamics leads to a large IMF multiplicity
at the end of the dynamical stage~\cite{NMD,Iwamoto},
the final yields are not necessarily that large~\cite{Iwamoto}.
On the other hand,
the quantum Langevin treatment leads to fragments
having a relatively small degree of excitation and, consequently,
a larger proportion of them survive the statistical decay.

The starting point for the present work
was the fact that wave packet molecular dynamics
displays classical statistical features,
an observation that has stimulated lively discussion
\cite{OR95,OR96a,eQMD,OR93,QMD-Stat,SF,OH}.
In our previous work~\cite{OR96a},
we demonstrated that when the mean energy is fixed
molecular dynamics in a common harmonic oscillator potential
may give the same observation with and without the quantal Langevin force.
However,
this coincidence depends strongly on the well-bound nature of the system
and the calculated results in the present paper suggest that 
additional fluctuations are necessary to achieve a satisfactory
molecular dynamics description of fragmentation processes.
A similar conclusion has been reached by Ono and Horiuchi~\cite{AMD-V}.
They included additional fluctuations to mimic the Vlasov dynamics in \AMD,
and also performed a constrained distortion of the wave function
at each point in time.
The combined effect of this approach seems to be rather similar
to the two main features of the present quantal Langevin model.

Before it is possible to quantitatively assess the role played
by the quantum nature of the system,
it is necessary to resolve a number of important ambiguities
in the quantal Langevin dynamics.
For example, the mobility tensor adopted in this present work
is based more on physical intuition than on first principles.
Another ambiguity concerns the residual two-body collisions,
since the quantal Langevin term
causes the level density of the final state to differ from the classical one.

However, taken at face value,
our results for Au+Au collisions indicate that the inclusion of the quantum
fluctuations in the wavepacket dynamics leads to a significant increase
in the production of massive fragments with low excitation.
Although the experimental data are not yet reproduced quantitatively,
the improvement is significant over the results obtained with the usual
treatments in which the quantum fluctuations are ignored.
This general qualitative result suggests that the underlying quantum nature
of the nuclear many-body system may indeed play a significant role
in fragmentation reactions.

\ack{
The authors are grateful to
Dr.\ M.B.\ Tsang for supplying us with the experimental data.
This work was supported in part by
the Grant-in-Aid for Scientific Research (No.\ 06740193)
from the Ministry of Education, Science and Culture, Japan,
and by the Director, Office of Energy Research,
Office of High Energy and Nuclear Physics,
Nuclear Physics Division of the U.S. Department of Energy
under Contract No.\ DE-AC03-76SF00098.
One of the authors (AO) also thanks the Ministry of Education,
Science and Culture, Japan, for the Overseas Research Fellowship granted.
}


\end{document}